\documentclass[12pt]{article}
\usepackage{amsmath}
\usepackage{amsthm}
\usepackage{amsfonts}
\usepackage[dvips]{epsfig}
\usepackage{graphicx}
\usepackage{amssymb}
\usepackage{cancel}
\usepackage{pstricks} 
\usepackage{lscape}
\usepackage{enumitem}
\usepackage{subcaption}

\usepackage[sort&compress,numbers, merge]{natbib}

\DeclareFontFamily{U}{mathx}{\hyphenchar\font45}
\DeclareFontShape{U}{mathx}{m}{n}{<-> mathx10}{}
\DeclareSymbolFont{mathx}{U}{mathx}{m}{n}

\newcommand{\beq}{\begin{equation}}
\newcommand{\eeq}{\end{equation}}

\makeatletter
\newlength{\apb@width}
\newcommand{\autoparbox}[2][c]{\settowidth{\apb@width}{#2}\parbox[#1]{\apb@width}{#2}}

\newcommand{\Cen}[2]{%
  \ifmeasuring@
    #2%
  \else
    \makebox[\ifcase\expandafter #1\maxcolumn@widths\fi]{$\displaystyle#2$}%
  \fi
}
\makeatother

\definecolor{Orange}{cmyk}{0,0.61,0.87,0}
\definecolor{JungleGreen}{cmyk}{0.99,0,0.52,0}
\definecolor{OliveGreen}{cmyk}{0.64,0,0.95,0.40}
\definecolor{Brown}{cmyk}{0,0.81,1,0.60}
\definecolor{RoyalBlue}{cmyk}{0.71,0.53,0,0.12}

\allowdisplaybreaks[1]

\usepackage[colorlinks=true, linkcolor=OliveGreen, citecolor=RoyalBlue,
urlcolor=RoyalBlue]{hyperref}

\begin{document}

\vspace{-0.2in}
\begin{flushright}
{\tt KCL-PH-TH/2020-10}, {\tt CERN-TH-2020-031}  \\
{\tt UT-20-03, ACT-01-20, MI-TH-206} \\
{\tt UMN-TH-3914/20, FTPI-MINN-20/04} \\
{\tt IFT-UAM/CSIC-20-36}
\end{flushright}

\vspace{0.05cm}
\begin{center}
{\bf \LARGE Proton Decay: Flipped vs Unflipped SU(5)}
\end{center}
\vspace{0.05cm}

\begin{center}{
{\bf John~Ellis}$^{a}$,
{\bf Marcos~A.~G.~Garcia}$^{b}$,
{\bf Natsumi Nagata}$^{c}$, \\[0.1cm]
{\bf Dimitri~V.~Nanopoulos}$^{d}$ and
{\bf Keith~A.~Olive}$^{e}$
}
\end{center}

\begin{center}
{\em $^a$Theoretical Particle Physics and Cosmology Group, Department of
  Physics, King's~College~London, London WC2R 2LS, United Kingdom;\\
Theoretical Physics Department, CERN, CH-1211 Geneva 23,
  Switzerland;\\
National Institute of Chemical Physics and Biophysics, R\"{a}vala 10, 10143 Tallinn, Estonia}\\[0.2cm]
  {\em $^b$Instituto de F\'isica Te\'orica (IFT) UAM-CSIC, Campus de Cantoblanco, 28049, Madrid, Spain}\\[0.2cm] 
  {\em $^c$Department of Physics, University of Tokyo, Bunkyo-ku, Tokyo
 113--0033, Japan}\\[0.2cm] 
{\em $^d$George P. and Cynthia W. Mitchell Institute for Fundamental
 Physics and Astronomy, Texas A\&M University, College Station, TX
 77843, USA;\\ 
 Astroparticle Physics Group, Houston Advanced Research Center (HARC),
 \\ Mitchell Campus, Woodlands, TX 77381, USA;\\ 
Academy of Athens, Division of Natural Sciences,
Athens 10679, Greece}\\[0.2cm]
{\em $^e$William I. Fine Theoretical Physics Institute, School of
 Physics and Astronomy, University of Minnesota, Minneapolis, MN 55455,
 USA}
 
 \end{center}

\vspace{0.5cm}
\centerline{\bf ABSTRACT}
\vspace{0.2cm}

{\small 
We analyze nucleon decay modes in a no-scale supersymmetric
flipped SU(5) GUT model, and contrast them with the predictions for
proton decays via dimension-6 operators in 
a standard unflipped supersymmetric SU(5) GUT model. We find that these
GUT models make very different predictions for the ratios
$\Gamma(p \to \pi^0 \mu^+)/\Gamma(p \to \pi^0 e^+)$, 
$\Gamma(p \to \pi^+ \bar \nu)/\Gamma(p \to \pi^0 e^+)$,
$\Gamma(p \to K^0 e^+)/\Gamma(p \to \pi^0 e^+)$ and
$\Gamma(p \to K^0 \mu^+)/\Gamma(p \to \pi^0 \mu^+)$, and
that predictions for the ratios
$\Gamma(p \to \pi^0 \mu^+)/\Gamma(p \to \pi^0 e^+)$ and 
$\Gamma(p \to \pi^+ \bar \nu)/\Gamma(p \to \pi^0 e^+)$ also
differ in variants
of the flipped SU(5) model with normal- or inverse-ordered
light neutrino masses. Upcoming large neutrino experiments may have interesting opportunities to explore both GUT and flavour physics in proton and neutron decays.
}

\vspace{0.5in}

\begin{flushleft}
March 2020
\end{flushleft}
\medskip
\noindent

\newpage

\section{Introduction}

The advent of a new generation of high-mass
underground neutrino detectors---JUNO~\cite{JUNO}, DUNE~\cite{DUNECDR} and
Hyper-Kamiokande~\cite{HKTDR}---will also open up new prospects
for searches for proton (and neutron) decays into
an array of channels with sensitivities an order of
magnitude beyond current experiments. This motivates
a re-evaluation of possible nucleon decay modes in
different grand unified theories (GUTs), 
and analyses of specific signatures
that may discriminate between the different models.
A well-known example is the distinction that can be
drawn between the minimal nonsupersymmetric SU(5) GUT
\cite{Georgi:1974sy}---in which the most characteristic proton decay mode is 
expected to be $p \to \pi^0 e^+$ induced by dimension-6 operators---and the minimal
supersymmetric SU(5) GUT~\cite{Dimopoulos:1981zb}---in
which the dominant decay mode is expected to be
$p \to K^+ \bar \nu$~\cite{Knu} induced by dimension-5 operators~\cite{Sakai:1981pk}. The prospective sensitivities
of the new generation of neutrino detectors to these
decay modes has been documented~\cite{JUNO,DUNECDR,HKTDR}, and the rate for
$p \to K^+ \bar \nu$ in the minimal supersymmetric 
SU(5) GUT has recently been re-evaluated, including 
an assessment of the uncertainties in the lifetime
estimate~\cite{Ellis:2019fwf}.

As is well known, the difference between the dominant nucleon decays
in the minimal supersymmetric and non-supersymmetric versions
of SU(5) is linked to the difference between their respective
decay mechanisms. Proton decay in minimal non-supersymmetric SU(5) 
is mediated by dimension-6 operators~\cite{Weinberg:1979sa}, 
whereas in minimal supersymmetric SU(5) $p \to K^+ \bar \nu$ is 
mediated by dimension-5 operators~\cite{Sakai:1981pk}.
The rate for dimension-5 proton decay is high enough to put
pressure on minimal supersymmetric SU(5) \cite{Goto:1998qg, mp}, though this problem is
mitigated by the higher sparticle masses \cite{Hisano:2013exa, Nagata:2013sba, Nagata:2013ive,
evno, eelnos, eemno, Evans:2019oyw, Ellis:2019fwf, McKeen:2013dma} now required by fruitless
LHC searches~\cite{nosusy, nosusy2}. Nevertheless, this issue has added to the motivations
for considering the supersymmetric flipped SU(5) GUT~\cite{Barr,DKN,flipped2,AEHN}, in which an economical
missing-partner mechanism \cite{flipped2, Masiero:1982fe, Hisano:1994fn} suppresses dimension-6 proton decay. This model
is also of interest because it can easily be accommodated within
string theory~\cite{AEHN, cehnt}, and a unified cosmological scenario for inflation,
dark matter, neutrino masses and baryogenesis has been constructed~\cite{egnno2, egnno3, egnno4, egnno5} in the
combined framework of flipped SU(5) and string-motivated \cite{Witten} no-scale supergravity~\cite{no-scale,ekn2,LN}.

The dominant final states for proton decay in supersymmetric
flipped SU(5) are not expected to contain strange particles,
with many of the favoured decay modes expected to be
similar to those in minimal supersymmetric SU(5),
including $p \to \pi^0 e^+$ and $\pi^+ \bar \nu$~\cite{Ellis:1988tx}. It is
therefore important to assemble a kit of diagnostic tools
that the upcoming experiments can use to discriminate
between the flipped and unflipped SU(5) GUT models.~\footnote{See~\cite{2002.11413} for proposed diagnostic tools for other GUT models.}
This issue has been discussed previously~\cite{Ellis:1993ks, Ellis:1995at, Ellis:2002vk, Dorsner:2004xx, Li:2010dp}, and the purpose
of this paper is to update the available diagnostic kit
in the framework of the unified cosmological framework
that we have proposed previously~\cite{egnno2, egnno3, egnno4, egnno5}, stressing the
connection between the flavour structure of nucleon decay
operators and the pattern of mixing between neutrinos
and their mass ordering.

We identify two primary proton decay signatures of 
the no-scale flipped SU(5) model~\cite{egnno2, egnno3, egnno4, egnno5} that may also cast light
on the mass-ordering of light neutrinos. One signature
is the ratio $\Gamma (p\to \pi^0 \mu^+)/
\Gamma (p \to \pi^0 e^+)$, and the other is
$\Gamma (p\to \pi^+ \bar{\nu})/\Gamma (p \to \pi^0 e^+)$.~\footnote{Here and subsequently, the sum
over the three light neutrino species is to be understood.}
In minimal SU(5) one expects $\Gamma (p\to \pi^0 \mu^+)/
\Gamma (p \to \pi^0 e^+) \sim 0.008$, whereas this ratio is
$\sim 0.1$ in flipped SU(5) with normally-ordered (NO)
light neutrinos and $\sim 23$ with inversely-ordered (IO)
neutrinos. In the case of $\Gamma (p\to \pi^+ \bar{\nu})/
\Gamma (p \to \pi^0 e^+)$, the IO flipped SU(5) model
predicts a ratio $\sim 95$ and the NO model predicts a 
ratio $\sim 3.2$, whereas the minimal SU(5) model allows
values as low as 0.4. In addition to these headline
signatures, we also find that the ratio $\Gamma (p\to K^0 e^+)/\Gamma (p\to \pi^0 e^+)$ would be larger in flipped
SU(5) than in minimal SU(5), $\sim 0.02$ vs $\sim 0.003$,
whereas the ratio $\Gamma (p\to K^0 \mu^+)/\Gamma (p\to \pi^0 \mu^+) \sim 0.02$
in the flipped SU(5) model, as opposed to $\sim 17$ in minimal SU(5).
It is clear therefore, that measurements of proton decay in more than one final state could discriminate between  underlying GUT models, and we show that searches for neutron decays may also play an important role.

The outline of this paper is the following. In Section~\ref{sec:model} we review relevant
features of the no-scale flipped SU(5) GUT model, and in Section~\ref{sec:protondecay} we
study proton (and some neutron) decay modes in this model, giving expressions in terms of 
the relevant hadronic matrix elements and discussing their uncertainties.
The corresponding expressions in unflipped SU(5) are discussed in
Section~\ref{sec:dim6protondecay}. 
In Section~\ref{sec:results} we present predictions for ratios of proton decay rates in
the flipped and unflipped SU(5) GUTs, and we review our conclusions and discuss future prospects
in Section~\ref{sec:conclusion}.

\section{The No-Scale Flipped SU(5) Model}
\label{sec:model}

In the no-scale flipped ${\rm SU}(5) \times {\rm U}(1)$ GUT model
\cite{Barr,DKN,flipped2,AEHN, egnno2, egnno3, egnno4, egnno5}, the three
generations of the minimal supersymmetric extension of the Standard Model (MSSM) matter
fields are embedded, together with three right-handed singlet neutrino chiral superfields,
into three sets of $\mathbf{10}$, $\bar{\mathbf{5}}$, and $\mathbf{1}$
representations of SU(5), which we denote by $F_i$, $\bar{f}_i$
and $\ell^c_i$, respectively, where $i=1,2,3$ is the generation
index. In units of $1/\sqrt{40}$, the U(1) charges of the $F_i$, $\bar{f}_i$ and $\ell^c_i$ are $+1$,
$-3$, and $+5$, respectively. The assignments
of the quantum numbers for the right-handed leptons, up-
and down-type quarks are ``flipped'' with respect to the standard SU(5)
assignments, giving the model its flippant name.
 
In addition to these matter fields, the minimal flipped SU(5) model contains a pair of
$\mathbf{10}$ and $\overline{\mathbf{10}}$ Higgs fields, $H$ and
$\bar{H}$, respectively, a pair of $\mathbf{5}$ and
$\overline{\mathbf{5}}$ Higgs fields, $h$ and $\bar{h}$, respectively,
and four singlet fields, $\phi_a$ ($a = 0, \dots, 3$). The vacuum
expectation values (VEVs) of the $H$ and $\bar{H}$ fields break the ${\rm
SU}(5) \times {\rm U}(1)$ gauge group down to the SM gauge group, and
subsequently the VEVs of the doublet Higgs fields $H_d$ and $H_u$, which
reside in $h$ and $\bar{h}$, respectively, break the $\mathrm{SU}(2)_L
\times \mathrm{U}(1)_Y$ gauge symmetry down to the U(1) of electromagnetism.  

The renormalizable superpotential in this model is given by
\begin{align} \notag
W &=  \lambda_1^{ij} F_iF_jh + \lambda_2^{ij} F_i\bar{f}_j\bar{h} +
 \lambda_3^{ij}\bar{f}_i\ell^c_j h + \lambda_4 HHh + \lambda_5
 \bar{H}\bar{H}\bar{h}\\ 
&+ \lambda_6^{ia} F_i\bar{H}\phi_a + \lambda_7^a h\bar{h}\phi_a
 + \lambda_8^{abc}\phi_a\phi_b\phi_c + \mu^{ab}\phi_a\phi_b\,.
\label{Wgen} 
\end{align}
We assume here that the model possesses an approximate $\mathbb{Z}_2$
symmetry, under which only the $H$ field is odd while the rest of the
fields are even. This symmetry is supposed to be violated by some
Planck-scale suppressed operators, which prevent the formation of domain
walls when the field $H$ acquires a VEV. This $\mathbb{Z}_2$
symmetry forbids some unwanted terms, such as $F_i H h$ and $\bar{f}_i H
\bar{h}$, which would cause baryon/lepton-number violation as well as
$R$-parity violation. The $\mathbb{Z}_2$ symmetry also forbids a
vector-like mass term for $H$ and $\bar{H}$, which is advantageous for
suppressing rapid proton decay induced by colour-triplet Higgs exchange. 

We embed the flipped SU(5) model in minimal $N = 1$ supergravity,
which we assume to have a K\"ahler potential of no-scale form~\cite{ekn2},
as is motivated by the low-energy structure of string theory~\cite{Witten}. In this case the potential $V$
has an $F$- and $D$-flat direction along a linear
combination of the singlet components in $H$ and $\bar{H}$. These fields
develop VEVs in this direction, as discussed in detail in Ref.~\cite{egnno3}. After $H$ and $\bar{H}$
acquire VEVs in this `flaton' direction, the coloured components in these fields form vector-like
multiplets with those in $h$ and $\bar{h}$ via the couplings $\lambda_4$
and $\lambda_5$ in (\ref{Wgen}). On the other hand, the electroweak doublets $H_d$ and
$H_u$ in $h$ and $\bar{h}$ do not acquire masses from the flaton
VEV---this is an economical realization of the missing-partner mechanism
\cite{flipped2} that solves naturally the
doublet-triplet splitting problem. 

As discussed in detail in Ref.~\cite{egnno2}, this model offers the possibility of
successful Starobinsky-like \cite{Staro} inflation, with one of the singlet fields,
$\phi_0$, playing the role of the inflaton \cite{ENO6}. For $\mu^{00} =
m_s/2$ and $\lambda_8^{000} = -m_s/(3\sqrt{3} M_P)$ in (\ref{Wgen}) with 
the inflaton mass $m_s \simeq 3
\times 10^{13}$~GeV and $M_P \equiv (8\pi G_N)^{-1/2}$ the reduced
Planck mass, the measured amplitude of the primordial power spectrum is
successfully reproduced and the tensor-to-scalar ratio $r \simeq 3
\times 10^{-3}$, well within the range allowed by the Planck results and other
data~\cite{Aghanim:2018eyx}. This prediction can be tested in future CMB
experiments such as CMB-S4~\cite{Abazajian:2016yjj} and LiteBIRD~\cite{Hazumi:2019lys}. 
The predicted value of the tilt in the scalar
perturbation spectrum, $n_s$, is also within the range favoured by
Planck and other data at the 68\% CL~\cite{Aghanim:2018eyx}. 

As seen in Eq.~\eqref{Wgen}, the inflaton $\phi_0$ can couple to the
matter sector via the couplings $\lambda_6$ and $\lambda_7$. In
Ref.~\cite{egnno2}, two distinct cases, $\lambda_6^{i0} = 0$ (Scenario
A) or $\lambda_6^{i0} \neq 0$ (Scenario B), were studied. We
focus on Scenario B in this work. In this scenario, one of the three
singlet fields other than $\phi_0$, which we denote by $\phi_3$, does
not have the $\lambda_6$ coupling; i.e., $\lambda_6^{i3} = 0$, whereas
$\lambda_6^{ia} \neq 0$ for $i = 1,2,3$ and $a = 0,1,2$. We also assume
$\lambda_7^a = 0$ for $a = 0,1,2$. To realize this scenario, we
introduce a modified $R$-parity, under which the fields in this model
transform as 
\begin{align}
    F_i, \bar{f}_i, \ell_i^c, \phi_0, \phi_1, \phi_2
    &\to 
    -F_i, -\bar{f}_i, -\ell_i^c, -\phi_0, -\phi_1, -\phi_2 ~, \nonumber \\[3pt]
    H, \bar{H}, h, \bar{h}, \phi_3 &\to 
    H, \bar{H}, h, \bar{h}, \phi_3 ~.
\end{align}
We note that this modified $R$-parity is slightly violated by the
coupling $\lambda_8^{000}$. Nevertheless, since this
$R$-parity-violating effect is only very weakly transmitted to the
matter sector, the lifetime of the lightest supersymmetric particle
(LSP) is still much longer than the age of the Universe \cite{egnno3,
ENO8}, so the LSP can be a good dark matter candidate. We also note that
the singlet $\phi_3$ can acquire a VEV without spontaneously breaking
the modified $R$-parity. In this case, the coupling $\lambda_7^3$, which
is allowed by the modified $R$-parity, generates an effective $\mu$ term
for $h$ and $\bar h$, $\mu = \lambda_7^3 \langle \phi_3 \rangle$, just as
in the next-to-minimal supersymmetric extension of the SM.  

As discussed in detail in Refs.~\cite{egnno2, egnno3, egnno4, egnno5}, the
$\lambda_6$ coupling in this model controls i) inflaton decays and
reheating; ii) the gravitino production rate and therefore the
non-thermal abundance of the LSP; iii) neutrino masses; and iv) the
baryon asymmetry of the Universe via leptogenesis \cite{fy}. In
particular, we showed in Refs.~\cite{egnno4, egnno5} by scanning over possible
values of $\lambda_6$ that the observed values of neutrino masses,
the dark matter abundance, and baryon asymmetry can be explained
simultaneously, together with a soft supersymmetry-breaking scale in the multi-TeV range. In this paper, we study nucleon decays in the
scenario developed in Refs.~\cite{egnno2, egnno3, egnno4, egnno5}.

Without loss of generality, we adopt the basis where $\lambda_2^{ij}$
and $\mu^{ab}$ are real and diagonal. In this case, the MSSM matter
fields and right-handed neutrinos are embedded into the SU(5)
representations as in~\cite{Ellis:1993ks}:~\footnote{We use the basis 
in which $U_u = U_{u^c} = U_\phi = 1$, where these matrices are as defined
in Ref.~\cite{Ellis:1993ks}. Moreover, we have removed the overall phase
factor $U_6$ using the field redefinition of $F_i$ and $\bar{f}_i$ and
expressed the diagonal phase matrix $U_7$ as $(U_7)_{ij} =
e^{i\varphi_i} \delta_{ij}$. }
\begin{align}
 F_i & \ni \left\{Q_i, ~ V_{ij} e^{-i \varphi_j} d_j^c, ~ \left(U_{\nu^c}\right)_{ij}
 \nu^c_j \right\} ~, \nonumber \\
 \bar{f}_i & \ni \left\{
u_i^c ~, L_j \left(U_l\right)_{ji}
\right\} ~, \nonumber \\
 \ell^c_i &= \left(U_{l^c}\right)_{ij} e_j^c ~,
\label{eq:embedding}
\end{align}
where the $V_{ij}$ are the Cabibbo-Kobayashi-Maskawa (CKM) matrix elements,
$U_{\nu^c}$, $U_l$, and $U_{l^c}$ are unitary matrices, and the phase
factors $\varphi_i$ satisfy the condition $\sum_{i} \varphi_i = 0$~\cite{Ellis:1993ks}. The
components of the doublet fields $Q_i$ and $L_i$ are written as 
\begin{equation}
 Q_i = 
\begin{pmatrix}
 u_i \\ V_{ij} d_j
\end{pmatrix}
~, \qquad
L_i = 
\begin{pmatrix}
 (U_{\rm PMNS})_{ij} \nu_j \\ e_i
\end{pmatrix}
~,
\end{equation}
where $U_{\rm PMNS}$ is the Pontecorvo-Maki-Nakagawa-Sakata (PMNS)
matrix.~\footnote{We define the PMNS matrix as in
the Review of Particle Physics (RPP)~\cite{PDG}, and that $U_{\rm PMNS} = U^*_{\rm MNS}$ in the notation of Ref.~\cite{Ellis:1993ks}.} 

The diagonal components of $\lambda_2^{ij}$ and $\mu^{ab}$ ($a,b
=0,1,2$) are given by 
\begin{equation}
 \lambda_2 \simeq \frac{1}{\langle \bar{h}_0 \rangle}
{\rm diag}(m_u, m_c, m_t) ~, \qquad
 \mu = \frac{1}{2}{\rm diag} (m_s, \mu^1, \mu^2) ~,
 \label{eq:lam2andmu}
\end{equation}
where we take $m_s = 3\times 10^{13}$~GeV (see above).
In what follows we 
express these matrices as  $\lambda_2^{ij} = \lambda_2^i \delta^{ij}$
and $\mu^{ab} = \mu^a \delta^{ab}$. The first equation
in Eq.~\eqref{eq:lam2andmu} is only an approximate expression, since
in general renormalization-group effects and threshold corrections cause
$\lambda_2$ to deviate from the up-type Yukawa couplings at low energies. However, since these effects are at most ${\cal O}(10)\%$
and depend on the mass
spectrum of the theory, we neglect them in the following
analysis.

The neutrino/singlet-fermion mass matrix can be written as 
\begin{equation}
 {\cal L}_{\rm mass}=
-\frac{1}{2}
\left(
\begin{matrix}
{\nu}_{i} & {\nu}_{j}^c & \tilde{\phi_a}
\end{matrix}
\right) 
\begin{pmatrix}
 0 & \lambda_2^{i j}\langle \bar{h}_0\rangle & 0 \\
\lambda_2^{i j}\langle \bar{h}_0\rangle & 0 &
 \lambda_6^{j a}\langle \tilde{\nu}_{\bar{H}}^c\rangle \\
0 & \lambda_6^{j a}\langle \tilde{\nu}_{\bar{H}}^c\rangle & \mu^a
\end{pmatrix}
 \left(
\begin{matrix}
\nu_{i} \\ \nu_{j}^c \\ \tilde{\phi_a}
\end{matrix}
\right)
+ {\rm h.c.}~,
\end{equation}
where $i,j = 1,2,3$ and $a= 0,1,2$,
and $\tilde{\phi}_0$ corresponds to the fermionic superpartner of the
inflaton field $\phi_0$. The mass matrix of the right-handed neutrinos
is then obtained from a first seesaw mechanism:
\begin{equation}
 (m_{\nu^c})_{ij} = \sum_{a=0,1,2} \frac{\lambda_6^{ia} \lambda_6^{ja}}{\mu^a}
  \langle \tilde{\nu}_{\bar{H}}^c \rangle^2 ~,
\label{eq:mnuc}
\end{equation}
where $\langle \tilde{\nu}_{\bar{H}}^c \rangle$ denotes the VEV of the
$F$- and $D$-flat direction of the singlet components of $H$ and
$\bar{H}$: we take $\langle \tilde{\nu}_{\bar{H}}^c \rangle =
10^{16}$~GeV in the following analysis. We diagonalize the mass matrix
in Eq.~\eqref{eq:mnuc} using a unitary matrix $U_{\nu^c}$:
\begin{equation}
     m_{\nu^c}^D = U_{\nu^c}^T m_{\nu^c}
U_{\nu^c}  ~.
\label{eq:mnucdiagonalization}
\end{equation}
The light neutrino mass matrix is then obtained through a second
seesaw mechanism~\cite{Minkowski:1977sc,Georgi:1979dq}:
\begin{equation}
 (m_\nu)_{ij} = \sum_{k} \frac{\lambda_2^i \lambda_2^j (U_{\nu^c})_{ik}
  (U_{\nu^c})_{jk} \langle \bar{h}_0 \rangle^2 }{(m_{\nu^c}^D)_k} ~.
\label{eq:mnu}
\end{equation}
This mass matrix is diagonalised by a unitary matrix $U_\nu$, so that 
\begin{equation}
    m_\nu^D
= U_\nu^* m_\nu  U_\nu^\dagger ~.
\label{eq:mnudiagonalization}
\end{equation}
We note that, given a matrix
$\lambda^{ia}_6$, the eigenvalues of the $m_\nu$ and $m_{\nu^c}$
matrices, as well as the mixing matrices $U_{\nu^c}$ and $U_\nu$, are
uniquely determined as functions of $\mu^1$ and $\mu^2$  via
Eqs.~(\ref{eq:mnuc}--\ref{eq:mnu}). 
The PMNS matrix is given by $U_l$ in
Eq.~\eqref{eq:embedding} and $U_\nu$ in
Eq.~\eqref{eq:mnudiagonalization}:
\begin{equation}
    U_{\rm PMNS} = U_l^{*} U_\nu^T ~.
    \label{eq:pmns}
\end{equation}
Using the measured values of the PMNS matrix elements, we can
use this relation to
obtain $U_l$ from $U_\nu$. The matrix $U_l$ plays
an important role in determining the partial decay widths of proton
decay modes, as we will see in the subsequent Section.

\section{Nucleon Decay in Flipped SU(5)}
\label{sec:protondecay}

We are now ready to discuss nucleon decay in our model. In view of the suppression of the dimension-5
contribution mediated by coloured Higgs fields 
thanks to the missing-partner mechanism in the 
flipped SU(5) GUT~\cite{flipped2}, the main contribution to
nucleon decay is due to exchanges of SU(5) 
gauge bosons. 
The relevant gauge interaction terms are 
\begin{align}
 K_{\rm gauge}&=
\sqrt{2}g_5\bigl(
-\epsilon_{\alpha\beta}(u^c_a)^{\dagger}X^\alpha_a U_l^T L^\beta
+\epsilon^{abc}(Q^{a\alpha})^{\dagger} X^\alpha_bVP^\dagger {d}^c_c
+ \epsilon_{\alpha\beta}(\nu^c)^\dagger U_{\nu^c}^\dagger X^\alpha_aQ^{a\beta}
+{\rm h.c.}
\bigr)~,
\label{eq:gaugeintflipped}
\end{align}
where $g_5$ is the SU(5) gauge coupling constant, 
the $X_a^\alpha$ are the
SU(5) gauge vector superfields, 
$P_{ij} \equiv e^{i\varphi_i} \delta_{ij}$, 
$\alpha, \beta$ are SU(2)$_L$ indices, and $a,b,c$ are
SU(3)$_C$ indices. 

Below the GUT scale, the effects of SU(5) gauge boson exchanges are
in general described by the dimension-six effective operators
\begin{equation}
 {\cal L}_{6}^{\rm eff}
=C_{6(1)}^{ijkl}{\cal O}^{6(1)}_{ijkl}
+C_{6(2)}^{ijkl}{\cal O}^{6(2)}_{ijkl}
~,
\label{eq:l6eff}
\end{equation}
where 
\begin{align}
 {\cal O}^{6(1)}_{ijkl}&=\int d^2\theta d^2\bar{\theta}~
\epsilon_{abc}\epsilon_{\alpha\beta}
\bigl(u^{c\dagger}_i\bigr)^a
\bigl(d^{c\dagger} _j\bigr)^b
e^{-\frac{2}{3}g^\prime B}
\bigl(e^{2g_3G}Q_k^\alpha\bigr)^cL^\beta_l~,
 \\
{\cal O}^{6(2)}_{ijkl}&=\int d^2\theta d^2\bar{\theta}
\epsilon_{abc}\epsilon_{\alpha\beta}~
Q^{a\alpha}_iQ^{b\beta}_j
e^{\frac{2}{3}g^\prime B}
\bigl(e^{-2g_3G}u^{c\dagger} _k\bigr)^c
e^{c\dagger} _l~,
\end{align}
with $G$ and $B$ the SU(3)$_C$ and U(1)$_Y$ gauge vector superfields,
respectively, and $g_3$ and $g^\prime$ the corresponding gauge
couplings. In the unflipped SU(5) GUT both of the Wilson coefficients 
$C^{ijkl}_{6(1,2)}$ are non-zero, but in flipped SU(5) only $C^{ijkl}_{6(1)}$
is non-zero, and is given by~\footnote{However, although $C^{ijkl}_{6(2)}$ vanishes in flipped SU(5),
we retain it in the following formulae so that it can also be used for the
unflipped case.}
\begin{align}
 C^{ijkl}_{6(1)}&= \frac{g_5^2}{M_X^2} (U_l)_{li} V_{kj}^*
 e^{i\varphi_j} ~,
\label{eq:dim6gutmatchflipped}
\end{align}
where $M_X$ is the SU(5) gauge boson mass. The Wilson coefficients are run
down to low energy scales using the renormalisation group
equations. The renormalisation factors for $C^{ijkl}_{6(n)}$ $(n = 1,2)$
between the GUT scale and the electroweak scale, $A_{S_n}$, are evaluated at
the one-loop level~\footnote{The two-loop RGEs for these coefficients
above the SUSY-breaking scale are given in Ref.~\cite{Hisano:2013ege}.}
as \cite{Munoz:1986kq, Abbott:1980zj}: 
\begin{align}
 A_{S_1} &=
 \biggl[
\frac{\alpha_3(\mu_{\text{SUSY}})}{\alpha_3(\mu_{\rm GUT})}
\biggr]^{\frac{4}{9}}
\biggl[
\frac{\alpha_2(\mu_{\text{SUSY}})}{\alpha_2(\mu_{\rm GUT})}
\biggr]^{-\frac{3}{2}}
\biggl[
\frac{\alpha_1(\mu_{\text{SUSY}})}{\alpha_1(\mu_{\rm GUT})}
\biggr]^{-\frac{1}{18}}
\nonumber \\
&\times
\biggl[
\frac{\alpha_3(m_Z)}{\alpha_3(\mu_{\rm SUSY})}
\biggr]^{\frac{2}{7}}
\biggl[
\frac{\alpha_2(m_Z)}{\alpha_2(\mu_{\rm SUSY})}
\biggr]^{\frac{27}{38}}
\biggl[
\frac{\alpha_1(m_Z)}{\alpha_1(\mu_{\rm SUSY})}
\biggr]^{-\frac{11}{82}} ~, \nonumber \\
 A_{S_2} &=
 \biggl[
\frac{\alpha_3(\mu_{\text{SUSY}})}{\alpha_3(\mu_{\rm GUT})}
\biggr]^{\frac{4}{9}}
\biggl[
\frac{\alpha_2(\mu_{\text{SUSY}})}{\alpha_2(\mu_{\rm GUT})}
\biggr]^{-\frac{3}{2}}
\biggl[
\frac{\alpha_1(\mu_{\text{SUSY}})}{\alpha_1(\mu_{\rm GUT})}
\biggr]^{-\frac{23}{198}}
\nonumber \\
&\times
\biggl[
\frac{\alpha_3(m_Z)}{\alpha_3(\mu_{\rm SUSY})}
\biggr]^{\frac{2}{7}}
\biggl[
\frac{\alpha_2(m_Z)}{\alpha_2(\mu_{\rm SUSY})}
\biggr]^{\frac{27}{38}}
\biggl[
\frac{\alpha_1(m_Z)}{\alpha_1(\mu_{\rm SUSY})}
\biggr]^{-\frac{23}{82}} ~,
\end{align}
where $m_Z$, $\mu_{\rm SUSY}$, and $\mu_{\rm GUT}$ denote the $Z$-boson
mass, the SUSY scale and the GUT scale, respectively, and $\alpha_A
\equiv g_A^2/(4\pi)$ with $g_A$ ($A = 1,2,3$) the gauge coupling
constants of the SM gauge groups. We give the electroweak-scale matching
conditions for each decay mode in what follows. 
Below the electroweak
scale, we take into account the perturbative QCD renormalization
factor, which is computed in Ref.~\cite{Nihei:1994tx} at the two-loop
level: $A_L = 1.247$. We then calculate the partial decay widths of various proton
decay modes by using the corresponding hadronic matrix elements, for which
we use the results obtained from the QCD lattice simulation performed in
Ref.~\cite{AISS}. The relevant hadronic matrix elements are listed in
Table~\ref{tab:w0}.

\begin{table}
\begin{center}
\caption{\it hadronic matrix elements used in our analysis, which are
 taken from Ref.~\cite{AISS}. The statistical and systematic
 uncertainties are indicated by (...)(...). The subscripts $e$ and $\mu$ indicate
 that the matrix elements are evaluated at the corresponding lepton kinematic points.
}
\label{tab:w0}
\begin{tabular}{lc}

\hline
\hline
Matrix element & Value [GeV$^2$]\\
 \hline
{$\langle \pi^0|(ud)_Ru_L|p\rangle_e$~~~~} &$-0.131(4)(13)$ \\
{$\langle \pi^0|(ud)_Ru_L|p\rangle_\mu$~~~~} &$-0.118(3)(12)$ \\
{$\langle \pi^+|(ud)_Rd_L|p\rangle$} &$-0.186 (6)(18)$ \\
{$\langle K^0|(us)_Ru_L|p\rangle_e$}   &  0.103(3)(11)  \\
{$\langle K^0|(us)_Ru_L|p\rangle_\mu$}   &  0.099(2)(10)  \\
{$\langle K^+|(us)_Rd_L|p\rangle$}  &$-0.049(2)(5)$  \\
{$\langle K^+|(ud)_Rs_L|p\rangle$}  & $-0.134$(4)(14) \\
\hline\hline
\end{tabular}
\end{center}
\end{table}


In the following we summarise the partial decay widths for the proton decay modes that
we discuss in this paper, as well as two relevant neutron decay modes.~\footnote{We note that these partial decay
widths do not depend on the phases $\varphi_i$.}
\subsection*{\underline{$p \to \pi^0 e^+$}}

The relevant effective operators below the electroweak scale are 
\begin{align}
 {\cal L}(p\to \pi^0 l^+_i)
&=C_{RL}(udul_i)\bigl[\epsilon_{abc}(u_R^ad_R^b)(u_L^cl_{Li}^{})\bigr]
+C_{LR}(udul_i)\bigl[\epsilon_{abc}(u_L^ad_L^b)(u_R^cl_{Ri}^{})\bigr]
~,
\label{eq:lagptopil}
\end{align}
where 
\begin{align}
 C_{RL}(udul_i)&=C^{111i}_{6(1)}(m_Z)~, \nonumber \\
 C_{LR}(udul_i)&=V_{j1}\bigl[
C^{1j1i}_{6(2)}(m_Z)+C^{j11i}_{6(2)}(m_Z)
\bigr]~.
\end{align}
Note that, since $C^{ijkl}_{6(2)} = 0$ in flipped SU(5), the second term
in Eq.~\eqref{eq:lagptopil} vanishes for this model. 
The partial decay width can be expressed as follows in terms of these coefficients at the hadronic scale:
\begin{equation}
 \Gamma (p\to  \pi^0 l^+_i)=
\frac{m_p}{32\pi}\biggl(1-\frac{m_\pi^2}{m_p^2}\biggr)^2
\bigl[
\vert {\cal A}_L(p\to \pi^0 l^+_i) \vert^2+
\vert {\cal A}_R(p\to \pi^0 l^+_i) \vert^2
\bigr]~,
\label{eq:aptopil}
\end{equation}
where
\begin{align}
 {\cal A}_L(p\to \pi^0 l^+_i)&=
C_{RL}(udul_i)\langle \pi^0\vert (ud)_Ru_L\vert p\rangle
~,\nonumber \\
 {\cal A}_R(p\to \pi^0 l^+_i)&=
C_{LR}(udul_i)\langle \pi^0\vert (ud)_Ru_L\vert p\rangle
~.
\end{align}
Setting $i = 1$ in
Eq.~\eqref{eq:aptopil}, we obtain
\begin{align}
  \Gamma (p\to  \pi^0 e^+)_{\rm flipped}&=
\frac{g_5^4m_p |V_{ud}|^2 |(U_l)_{11}|^2
}{32\pi M_X^4}\biggl(1-\frac{m_\pi^2}{m_p^2}\biggr)^2
A_L^2 A_{S_1}^2 \left(\langle \pi^0\vert (ud)_Ru_L\vert
 p\rangle_{e}\right)^2~,
\label{eq:ptopiefl}
\end{align}
where $m_p$ and $m_\pi$ denote the masses of the proton and pion,
respectively, and here and subsequently the subscript on the hadronic matrix element indicates
that it is evaluated at the corresponding lepton kinematic point. 

From Eq.~\eqref{eq:ptopiefl}, we can readily compute the partial lifetime of 
the $p\to  \pi^0 e^+$ mode as 
\begin{align}
    \tau (p\to  \pi^0 e^+)_{\rm flipped}&\simeq 
    7.9 \times 10^{35} \times  |(U_l)_{11}|^{-2}
    \biggl(\frac{M_X}{10^{16}~{\rm GeV}}\biggr)^4
     \biggl(\frac{0.0378}{\alpha_5}\biggr)^2 ~{\rm yrs}~.
     \label{tp0ef}
\end{align}
We note that this tends to be longer than the lifetime predicted in unflipped SU(5) 
by a factor (see also Eq.~\eqref{eq:ptopie}) 
\begin{align}
    \frac{\tau (p\to  \pi^0 e^+)_{\rm flipped}}{\tau (p\to  \pi^0 e^+)_{\rm unflipped}}
    \simeq \frac{A_{S_1}^2 + (1+|V_{ud}|^2)^2 A_{S_2}^2} {A_{S_1}^2 |(U_l)_{11}|^2} 
    \simeq \frac{4.8}{|(U_l)_{11}|^2} ~,
\end{align}
as found in Refs.~\cite{Ellis:1988tx, Ellis:1993ks, mp, Ellis:2002vk}.~\footnote{Values of $(U_l)_{11}$
in specific flipped SU(5) GUT scenarios are discussed later: see Eqn.~(\ref{eq:ul11}).}

\subsection*{\underline{$p \to \pi^0 \mu^+ $}}

By using the effective Lagrangian in Eq.~\eqref{eq:lagptopil} and the rate in Eq.~\eqref{eq:aptopil} for $i =
2$, we have 
\begin{align}
  \Gamma (p\to  \pi^0 \mu^+)_{\rm flipped}&=
\frac{g_5^4m_p |V_{ud}|^2 |(U_l)_{21}|^2
}{32\pi M_X^4}\biggl(1-\frac{m_\pi^2}{m_p^2}\biggr)^2
A_L^2 A_{S_1}^2 \left(\langle \pi^0\vert (ud)_Ru_L\vert
 p\rangle_{\mu}\right)^2~,
\label{eq:ptopimufl}
\end{align}
and the partial lifetime of 
the $p\to  \pi^0 \mu^+$ mode is 
\begin{align}
    \tau (p\to  \pi^0 \mu^+)_{\rm flipped}&\simeq 
    9.7 \times 10^{35} \times  |(U_l)_{21}|^{-2}
    \biggl(\frac{M_X}{10^{16}~{\rm GeV}}\biggr)^4
     \biggl(\frac{0.0378}{\alpha_5}\biggr)^2 ~{\rm yrs}~.
     \label{tp0mf}
\end{align}

\subsection*{\underline{$n \to \pi^- l^+ $}}

We note in passing that the rates of neutron decay modes 
that include a charged lepton can be obtained from $\Gamma (p\to  \pi^0 l^+_i)$ 
through SU(2) isospin relations: 
\begin{equation}
    \Gamma (n\to  \pi^- l^+_i) = 2\Gamma (p\to  \pi^0 l^+_i) ~,
    \label{npi-l+}
\end{equation}
which applies to both the flipped and unflipped SU(5) models.

\subsection*{\underline{$p \to \pi^+ \bar{\nu}_i$}}

The relevant effective Lagrangian term in this case is
\begin{align}
 {\cal L}(p\to \pi^+ \bar{\nu}_i)
&=C_{RL}(udd\nu_i)\bigl[\epsilon_{abc}(u_R^ad_R^b)(d_L^c\nu_{Li}^{})\bigr]
~,
\end{align}
with the following matching condition at the electroweak scale 
\begin{align}
 C_{RL}(udd\nu_i)&=-V_{j1}C^{11ji}_{6(1)}(m_Z)~.
\end{align}
The partial decay width is then computed as
\begin{equation}
 \Gamma (p\to  \pi^+ \bar{\nu}_i)=
\frac{m_p}{32\pi}\biggl(1-\frac{m_\pi^2}{m_p^2}\biggr)^2
\vert {\cal A}(p\to \pi^+ \bar{\nu}_i) \vert^2~,
\end{equation}
with
\begin{align}
 {\cal A}(p\to \pi^+ \bar{\nu}_i)&=
C_{RL}(udd\nu_i)\langle \pi^+\vert (ud)_Rd_L\vert p\rangle
~.
\end{align}
We then have
\begin{align}
 \Gamma (p\to  \pi^+ \bar{\nu}_i)_{\rm flipped}&=
\frac{g_5^4m_p |(U_l)_{i1}|^2}{32\pi M_X^4}\biggl(1-\frac{m_\pi^2}{m_p^2}\biggr)^2
A_L^2 {A}_{S_1}^2 \left(\langle \pi^+\vert
 (ud)_Rd_L\vert p\rangle\right)^2 
~.
\label{eq:ptopinufl}
\end{align}

\subsection*{\underline{$n \to \pi^0 \bar{\nu}_i $}}

There is a relation between the partial decay widths 
for $n \to \pi^0 \bar{\nu}_i$ and those of $p\to  \pi^+ \bar{\nu}_i$ 
given by isospin: 
\begin{equation}
    \Gamma (n \to \pi^0 \bar{\nu}_i) = \frac{1}{2} 
    \Gamma(p\to  \pi^+ \bar{\nu}_i) ~,
    \label{eq:ntopi0nui}
\end{equation}
which applies to both the flipped and unflipped SU(5) models.

\subsection*{\underline{$p \to K^0 e^+$}}

The effective interactions in this case are given by
\begin{align}
 {\cal L}(p\to K^0 l^+_i)
&=C_{RL}(usul_i)\bigl[\epsilon_{abc}(u_R^as_R^b)(u_L^cl_{Li}^{})\bigr]
+C_{LR}(usul_i)\bigl[\epsilon_{abc}(u_L^as_L^b)(u_R^cl_{Ri}^{})\bigr]
~,
\label{eq:lagptokl}
\end{align}
with
\begin{align}
 C_{RL}(usul_i)&=C^{121i}_{6(1)}(m_Z)~, \nonumber \\
 C_{LR}(usul_i)&=V_{j2}\bigl[C^{1j1i}_{6(2)}(m_Z)
+C^{j11i}_{6(2)}(m_Z)
\bigr]~.
\end{align}
We then obtain the partial decay width
\begin{equation}
 \Gamma (p\to  K^0 l^+_i)=
\frac{m_p}{32\pi}\biggl(1-\frac{m_K^2}{m_p^2}\biggr)^2
\bigl[
\vert {\cal A}_L(p\to K^0 l^+_i) \vert^2+
\vert {\cal A}_R(p\to K^0 l^+_i) \vert^2
\bigr]~,
\label{eq:gamptokefl}
\end{equation}
where $m_K$ is the kaon mass and 
\begin{align}
 {\cal A}_L(p\to K^0 l^+_i)&=
C_{RL}(usul_i)\langle K^0\vert (us)_Ru_L\vert p\rangle
~,\nonumber \\
 {\cal A}_R(p\to K^0 l^+_i)&=
C_{LR}(usul_i)\langle K^0\vert (us)_Ru_L\vert p\rangle
~.
\end{align}
In particular, for $i = 1$, we have
\begin{align}
  \Gamma (p\to  K^0 e^+)_{\rm flipped}&=
\frac{g_5^4m_p |V_{us}|^2 |(U_l)_{11}|^2
}{32\pi M_X^4}\biggl(1-\frac{m_K^2}{m_p^2}\biggr)^2
A_L^2 A_{S_1}^2 \left(\langle K^0\vert (us)_Ru_L\vert
 p\rangle_{e}\right)^2~.
\label{eq:ptokefl}
\end{align}

\subsection*{\underline{$p \to K^0 \mu^+$}}

With $i=2$ in Eq.~\eqref{eq:gamptokefl}, we have 
\begin{align}
  \Gamma (p\to  K^0 \mu^+)_{\rm flipped}&=
\frac{g_5^4m_p |V_{us}|^2 |(U_l)_{21}|^2
}{32\pi M_X^4}\biggl(1-\frac{m_K^2}{m_p^2}\biggr)^2
A_L^2 A_{S_1}^2 \left(\langle K^0\vert (us)_Ru_L\vert
 p\rangle_{\mu}\right)^2~.
\label{eq:ptokmufl}
\end{align}

\subsection*{\underline{$p \to K^+ \bar{\nu}_i$}}

The low-energy effective interactions for this decay mode is given by
\begin{align}
 {\cal L}(p\to K^+\bar{\nu}_i^{})
=&
C_{RL}(usd\nu_i)\bigl[\epsilon_{abc}(u_R^as_R^b)(d_L^c\nu_i^{})\bigr]
+C_{RL}(uds\nu_i)\bigl[\epsilon_{abc}(u_R^ad_R^b)(s_L^c\nu_i^{})\bigr]
~,
\end{align}
with
\begin{align}
 C_{RL}(usd\nu_i)&=-V_{j1}C^{12ji}_{6(1)}(m_Z)~,\nonumber \\
 C_{RL}(uds\nu_i)&=-V_{j2}C^{11ji}_{6(1)}(m_Z)~.
\end{align}
We note that the unitarity of the CKM matrix leads to 
\begin{equation}
 V_{j1}C^{12ji}_{6(1)} = V_{j2}C^{11ji}_{6(1)} = 0~,
\end{equation}
in the case of flipped SU(5). As a result, we have 
\begin{align}
 \Gamma (p\to  K^+ \bar{\nu}_i)&=
0 ~,
\label{eq:ptoknufl}
\end{align}
as found in Ref.~\cite{Ellis:1993ks}.

\section{Dimension-Six Proton Decay in Unflipped SU(5)}
\label{sec:dim6protondecay}

In this Section we review briefly the proton decay calculation in unflipped SU(5), assuming
that proton decay is dominantly induced by dimension-6 SU(5) gauge boson
exchange, i.e., that the dimension-5 contribution of colour-triplet Higgs exchange is negligible. 
This assumption is valid, e.g., when the sfermion masses are
sufficiently large, i.e., $\gtrsim 100$~TeV \cite{Hisano:2013exa,
Nagata:2013sba, Nagata:2013ive, evno, eelnos, eemno, Evans:2019oyw, Ellis:2019fwf}
or if a suitable missing-partner mechanism is invoked~\cite{Masiero:1982fe, Hisano:1994fn}. For more
detailed discussions of the calculation of proton decay induced by
SU(5) gauge boson exchange in unflipped SU(5), see
Refs.~\cite{Hisano:1992jj, Hisano:2012wq, Nagata:2013ive, Evans:2019oyw, Ellis:2019fwf}.

In unflipped SU(5), the Wilson coefficients of the effective
operators in Eq.~\eqref{eq:l6eff} are given by
\begin{align}
 C^{ijkl}_{6(1)}&=-\frac{g_5^2}{M_X^2}e^{i\varphi_i}\delta^{ik}
\delta^{jl}~,
\nonumber \\
 C^{ijkl}_{6(2)}&=-\frac{g_5^2}{M_X^2}e^{i\varphi_i}
\delta^{ik}(V^*)^{jl}~.
\label{eq:dim6gutmatch}
\end{align}
The rest of the calculation is exactly the same as before, so we just
summarize the resultant expression for each partial decay width.

\subsection*{\underline{$p \to  \pi^0 e^+$}}

\begin{align}
 \Gamma (p\to  \pi^0 e^+)&=
\frac{g_5^4m_p}{32\pi M_X^4}\biggl(1-\frac{m_\pi^2}{m_p^2}\biggr)^2
A_L^2 \left(\langle \pi^0\vert (ud)_Ru_L\vert p\rangle_e\right)^2
\bigl[
{A}_{S_1}^2 + (1+|V_{ud}|^2)^2 {A}_{S_2}^2
\bigr]~.
\label{eq:ptopie}
\end{align}

\subsection*{\underline{$p \to \pi^0 \mu^+ $}}

\begin{align}
 \Gamma (p\to  \pi^0 \mu^+)&=
\frac{g_5^4m_p}{32\pi M_X^4}\biggl(1-\frac{m_\pi^2}{m_p^2}\biggr)^2
A_L^2 A_{S_2}^2 \left(\langle \pi^0\vert (ud)_Ru_L\vert p\rangle_\mu\right)^2
\bigl[
 \left|V_{ud}^{} V_{us}^*\right|^2 
\bigr]~.
\label{eq:ptopimu}
\end{align}

\subsection*{\underline{$p \to \pi^+ \bar{\nu}$}}

\begin{align}
 \Gamma (p\to  \pi^+ \bar{\nu}_i)&=
\frac{g_5^4m_p |V_{ud}|^2}{32\pi M_X^4}\biggl(1-\frac{m_\pi^2}{m_p^2}\biggr)^2
A_L^2 {A}_{S_1}^2 \delta^{1i} \left(\langle \pi^+\vert
 (ud)_Rd_L\vert p\rangle\right)^2 
~.
\label{eq:ptopinu}
\end{align}

\subsection*{\underline{$p \to K^0 e^+$}}

\begin{align}
 \Gamma (p\to  K^0 e^+)&=
\frac{g_5^4m_p}{32\pi M_X^4}\biggl(1-\frac{m_K^2}{m_p^2}\biggr)^2
A_L^2 A_{S_2}^2 \left(\langle K^0\vert (us)_Ru_L\vert p\rangle_e\right)^2
\bigl[
 \left|V_{ud}^{} V_{us}^*\right|^2 
\bigr]~.
\label{eq:ptoke}
\end{align}

\subsection*{\underline{$p \to K^0 \mu^+$}}

\begin{align}
 \Gamma (p\to  K^0 \mu^+)&=
\frac{g_5^4m_p}{32\pi M_X^4}\biggl(1-\frac{m_K^2}{m_p^2}\biggr)^2
A_L^2 \left(\langle K^0\vert (us)_Ru_L\vert p\rangle_\mu\right)^2
\bigl[
{A}_{S_1}^2 + (1+|V_{us}|^2)^2 {A}_{S_2}^2
\bigr]~.
\label{eq:ptokmu}
\end{align}

\subsection*{\underline{$p \to K^+ \bar{\nu}$}}

\begin{align}
 \Gamma (p\to  K^+ \bar{\nu}_i)&=
\frac{g_5^4m_p}{32\pi M_X^4}\biggl(1-\frac{m_K^2}{m_p^2}\biggr)^2
A_L^2 {A}_{S_1}^2 
\nonumber \\
&\times
\biggl[
\delta^{1i} |V_{us}|^2\left(\langle K^+\vert
 (ud)_Rs_L\vert p\rangle\right)^2 
+ \delta^{2i} |V_{ud}|^2\left(\langle K^+\vert
 (us)_Rd_L\vert p\rangle\right)^2 
\biggr]
~.
\end{align}

\section{Comparison of Proton Decay Rates in 
Flipped and Unflipped SU(5)}
\label{sec:results}

As we now discuss, the predictions for proton decay 
branching fractions in the flipped
SU(5) GUT model are different from those generated by dimension-6 operators in the standard unflipped SU(5) GUT,~\footnote{We assume here that the contributions of dimension-5 operators are suppressed, either by large sparticle and/or triplet Higgs masses, or by some missing-partner mechanism.} which may enable future experiments to distinguish these two GUT scenarios. To this end, we focus on the following five quantities
and compare the predictions for them in flipped and
unflipped SU(5) GUTs:
\begin{itemize}
 \item[i)] $\Gamma (p\to \pi^0 \mu^+)/\Gamma (p\to \pi^0 e^+)$ ~,
 \item[ii)] $\sum_i \Gamma (p\to \pi^+ \bar{\nu}_i)/\Gamma (p\to \pi^0
	  e^+)$ ~,
 \item[iii)] $\Gamma (p\to K^0 e^+)/\Gamma (p\to \pi^0 e^+)$ ~,
 \item[iv)] $\Gamma (p\to K^0 \mu^+)/\Gamma (p\to \pi^0 \mu^+)$ ~,
 \item[v)] $p \to K^+ \bar{\nu}$ ~.
\end{itemize}

\subsection{$\Gamma (p\to \pi^0 \mu^+)/\Gamma (p\to \pi^0 e^+)$}

From Eqs.~\eqref{eq:ptopiefl} and \eqref{eq:ptopimufl}, we find that
this ratio in the flipped SU(5) is given by
\begin{align}
 \frac{\Gamma (p\to  \pi^0 \mu^+)_{\rm
 flipped}}{\Gamma (p\to  \pi^0 e^+)_{\rm flipped}}
= \frac{\left(\langle \pi^0\vert (ud)_Ru_L\vert p\rangle_\mu\right)^2 
 |(U_l)_{21}|^2 }{
\left(\langle \pi^0\vert (ud)_Ru_L\vert p\rangle_e\right)^2
|(U_l)_{11}|^2
} ~.
\label{eq:ptopimuovefl}
\end{align}
We see that this ratio depends on the unitary matrix $U_l$, which is
determined from $U_\nu$ and the PMNS matrix $U_{\rm PMNS}$ via
Eq.~\eqref{eq:pmns}. We also note that by taking the ratio between the
two partial decay widths $\Gamma (p\to \pi^0 \mu^+)$ and $\Gamma (p\to
\pi^0 e^+)$, many of the factors in these quantities such as the SU(5)
gauge boson mass, $M_X$, the SU(5) gauge coupling constant, $g_5$, and the
renormalization factors, $A_L$ and $A_{S_1}$, are cancelled,
which makes
the prediction for this ratio rather robust.

In unflipped SU(5), on the other hand, we obtain (see
Eqs.~\eqref{eq:ptopie} and \eqref{eq:ptopimu}):
\begin{align}
\frac{\Gamma (p\to  \pi^0 \mu^+)_{\rm unflipped}}{\Gamma (p\to  \pi^0
 e^+)_{\rm unflipped}}
=
\frac{\left(\langle \pi^0\vert (ud)_Ru_L\vert p\rangle_\mu\right)^2 }
{\left(\langle \pi^0\vert (ud)_Ru_L\vert p\rangle_e\right)^2}
 \frac{ \left|V_{ud}^{} V_{us}^*\right|^2 }{
\bigl[
R_A^2 + (1+|V_{ud}|^2)^2 
\bigr]
} ~,
\end{align}
where 
\begin{equation}
 R_A \equiv \frac{A_{S_1}}{A_{S_2}} 
= 
\biggl[
\frac{\alpha_1(\mu_{\text{SUSY}})}{\alpha_1(\mu_{\rm GUT})}
\biggr]^{\frac{2}{33}}
\biggl[
\frac{\alpha_1(m_Z)}{\alpha_1(\mu_{\rm SUSY})}
\biggr]^{\frac{6}{41}} ~.
\end{equation}
We find $R_A \simeq 1$ in a typical supersymmetric mass spectrum, and for
$R_A = 1$ we have:~\footnote{This result is consistent with the formula given in
Ref.~\cite{Ellis:1979hy} 
\begin{equation}
 \biggl(\frac{\Gamma (p\to \mu^+ + X)}{\Gamma (p\to e^+ +
  X)}\biggr)_{X~\text{nonstrange}} 
= \frac{\sin^2 \theta_c \cos^2 \theta_c}{(1+\cos^2 \theta_c)^2 + 1} 
\simeq 0.01
~,
\end{equation}
where $\theta_c$ is the Cabibbo angle: $\sin\theta_c \simeq 0.2245$. 
}
\begin{equation}
 \frac{\Gamma (p\to  \pi^0 \mu^+)_{\rm unflipped}}{\Gamma (p\to  \pi^0
 e^+)_{\rm unflipped}}
\simeq 0.008 ~.
\end{equation}
Hence, the branching fraction of the muon mode is predicted
to be smaller than that of the electron mode by approximately two orders
of magnitude in the unflipped SU(5) GUT. This prediction is again rather
robust: the uncertainty is ${\cal O}(10)$\%, which mainly comes from the
errors in the hadronic matrix elements. We note also that the contribution
of the color-triplet Higgs exchange to these decay modes in
supersymmetric SU(5) is suppressed
by small Yukawa couplings, and thus is negligible unless there is
flavor violation in the sfermion mass matrices \cite{Nagata:2013sba}. 

To determine the predicted value of the ratio in flipped SU(5) given by Eq.~\eqref{eq:ptopimuovefl},
we perform a parameter scan similar to that in Refs.~\cite{egnno4,
egnno5}. We first write the Yukawa matrix $\lambda_6$ in the form
\begin{equation}
  \lambda_6 = r_6 M_6 ~,  
\end{equation}
where $r_6$ is a real constant, which plays a role of a scale factor,
and $M_6$ is a generic complex $3\times 3$ matrix. We then scan 
$r_6$ with a logarithmic distribution over the range $(10^{-4},
1)$ choosing a total of 1000 values. For each value of $r_6$, we
generate 1000 random complex $3\times 3$ matrices $M_6$ with each
component taking a value of ${\cal O}(1)$. 

As discussed in Refs.~\cite{egnno4, egnno5}, for each $3\times 3$ matrix
$\lambda_6$, the eigenvalues of the $m_\nu$ and $m_{\nu^c}$ matrices and
the mixing matrices $U_{\nu^c}$ and $U_\nu$ are obtained as functions of
$\mu^1$ and $\mu^2$ in Eq.~\eqref{eq:lam2andmu}. We then determine these
two $\mu$ parameters by requiring that the observed values of the
squared mass differences, $\Delta m_{21}^2 \equiv m_2^2 - m_1^2$ and
$\Delta m_{3\ell }^2 \equiv m_3^2 - m_\ell^2$, are reproduced within the
experimental uncertainties, where $\ell = 1$ for the NO case and $\ell = 2$ for the IO case. For the
experimental input, we use the results from $\nu$-fit 4.0 given in Ref.~\cite{nufit}. 
By using $U_\nu$ determined in this manner, we then compute the matrix
$U_l$ using the relation \eqref{eq:pmns}. We parametrise the PMNS matrix elements
following the RPP convention \cite{PDG}:
\begin{equation}
 U_{\rm PMNS} = 
\begin{pmatrix}
 c_{12} c_{13} & s_{12} c_{13} & s_{13} e^{-i\delta} \\
 -s_{12} c_{23} -c_{12} s_{23} s_{13} e^{i\delta}
& c_{12} c_{23} -s_{12} s_{23} s_{13} e^{i\delta}
& s_{23} c_{13}\\
s_{12} s_{23} -c_{12} c_{23} s_{13} e^{i\delta}
& -c_{12} s_{23} -s_{12} c_{23} s_{13} e^{i\delta}
& c_{23} c_{13}
\end{pmatrix}
\begin{pmatrix}
 1 & 0 & 0\\
 0 & e^{i\frac{\alpha_{2}}{2}} & 0\\
 0 & 0 & e^{i\frac{\alpha_{3}}{2}}
\end{pmatrix}
~,
\end{equation}
where $c_{ij} \equiv \cos \theta_{ij}$ and $s_{ij} \equiv \sin
\theta_{ij}$ with the mixing angles $\theta_{ij} = [0, \pi/2]$, the
Dirac CP phase $\delta \in [0, 2\pi]$, and the order $m_1<m_2$ is chosen
without loss of generality. Again we use the values obtained in
Ref.~\cite{nufit}  for $\theta_{12}$, $\theta_{23}$,
$\theta_{13}$, and $\delta$. As for the Majorana phases $\alpha_2$ and $\alpha_3$, we set
$\alpha_2 = \alpha_3 = 0$ in this analysis since, as we shall see below, the
result scarcely depends on these phases. 
We generate the same number of $\lambda_6$ matrices for each mass
ordering, and find solutions for 2399 and 180 matrix choices for
the NO and IO cases, respectively, out of a total of $10^6$ parameter
sets sampled. This difference indicates some preference for the NO case in
our model.

\begin{figure}[t]
\begin{center}
\includegraphics[height=75mm]{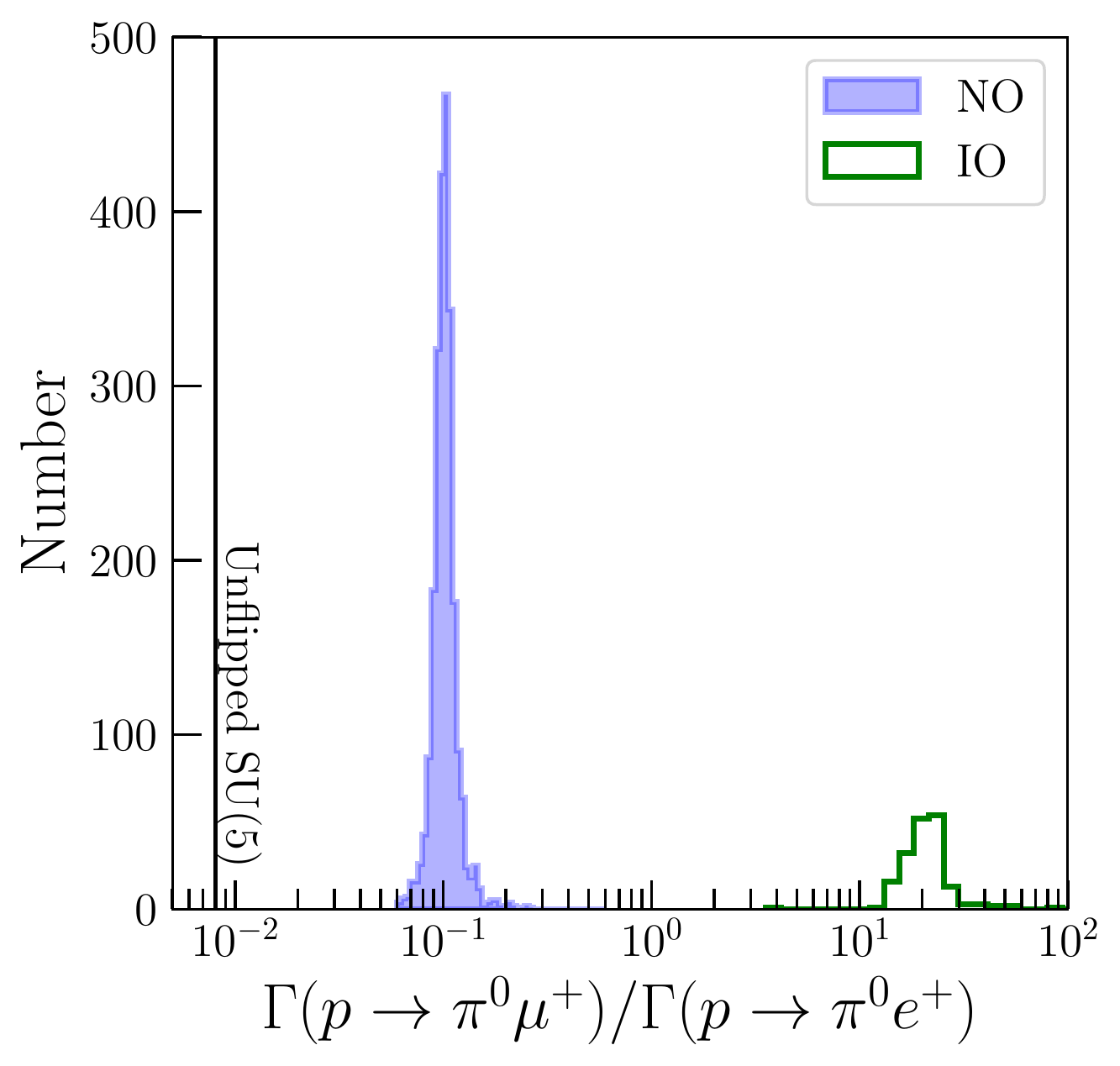}
\caption{\it Histograms of $\Gamma (p\to \pi^0 \mu^+)/\Gamma (p\to
 \pi^0 e^+)$ in the flipped SU(5) GUT model for the NO and IO cases in blue and
 green, respectively. The vertical line corresponds to the unflipped
 SU(5) prediction. }
\label{fig:rpimuove}
\end{center}
\end{figure}

In Fig.~\ref{fig:rpimuove} we display histograms of the ratio $\Gamma (p\to
\pi^0 \mu^+)/\Gamma (p\to \pi^0 e^+)$ in the NO and IO scenarios in blue
and green, respectively. The vertical black solid line represents the
predicted value in unflipped SU(5). As we see, the flipped SU(5) Model
predicts this ratio to be $\sim 0.10$ and $\sim 23$ for the NO and IO
cases, respectively. To understand the origin of these values, we first
note that, due to the hierarchical structure of $m_\nu$ in
Eq.~\eqref{eq:mnu}, $U_\nu$ has a simple form: 
\begin{equation}
 U_\nu \simeq 
\begin{pmatrix}
 1 & 0 & 0 \\
 0 & \cos \theta & - \sin \theta \\
 0 & \sin \theta & \cos \theta 
\end{pmatrix}
~,
\end{equation}
for NO, where $\sin \theta$ is found to be $\sim 0.38$, and 
\begin{equation}
 U_\nu \simeq 
\begin{pmatrix}
 0 & 1 & 0 \\
 0 & 0 & 1 \\
 1 & 0 & 0 
\end{pmatrix}
\begin{pmatrix}
 1 & 0 & 0 \\
 0 & 1/\sqrt{2} & - 1/\sqrt{2} \\
 0 & 1/\sqrt{2} & 1/\sqrt{2} 
\end{pmatrix}
~,
\end{equation}
for IO, where the first matrix in the right-hand side arranges the order
of the neutrino mass eigenvalues in accordance with the RPP
convention. The relevant matrix elements of $U_l = U_{\rm PMNS}^* U_\nu$
are then given by
\begin{align}
 (U_l)_{11}
&\simeq 
\begin{cases}
 (U_{\rm PMNS}^*)_{11} = c_{12} c_{13} &\qquad \text{NO} \\[5pt]
 (U_{\rm PMNS}^*)_{13} = s_{13} e^{i\delta - i\frac{\alpha_3}{2}} &\qquad \text{IO}
\end{cases}
~, \label{eq:ul11}\\[3pt]
 (U_l)_{21}
&\simeq
\begin{cases}
 (U_{\rm PMNS}^*)_{21} =  -s_{12} c_{23} -c_{12} s_{23} s_{13}
 e^{-i\delta} &\qquad \text{NO} \\[5pt]
 (U_{\rm PMNS}^*)_{23} = s_{23} c_{12} e^{-i\frac{\alpha_3}{2}} &\qquad \text{IO}
\end{cases}
~, \label{eq:ul21}
\end{align}
which leads to
\begin{align}
 \frac{\Gamma (p\to  \pi^0 \mu^+)_{\rm
 flipped}}{\Gamma (p\to  \pi^0 e^+)_{\rm flipped}}
\simeq  \frac{\left(\langle \pi^0\vert (ud)_Ru_L\vert p\rangle_\mu\right)^2  }{
\left(\langle \pi^0\vert (ud)_Ru_L\vert p\rangle_e\right)^2
} \frac{
 |s_{12} c_{23} +c_{12} s_{23} s_{13}
 e^{-i\delta}|^2 }{
(c_{12} c_{13})^2
} \simeq 0.10~,
\end{align}
for NO, and 
\begin{align}
 \frac{\Gamma (p\to  \pi^0 \mu^+)_{\rm
 flipped}}{\Gamma (p\to  \pi^0 e^+)_{\rm flipped}}
\simeq  \frac{\left(\langle \pi^0\vert (ud)_Ru_L\vert p\rangle_\mu\right)^2  }{
\left(\langle \pi^0\vert (ud)_Ru_L\vert p\rangle_e\right)^2
} \frac{
(s_{23} c_{12})^2 }{
s_{13}^2
} \simeq 22.9~,
\end{align}
for IO. These approximate estimates are in good agreement with the
results given in Fig.~\ref{fig:rpimuove}. We also note that these two
expressions do not depend on the unknown Majorana phases, $\alpha_2$ and
$\alpha_3$. As a consequence, although we have fixed these phases to be
zero in our analysis, we expect that the results in
Fig.~\ref{fig:rpimuove} will not be changed even if we take different
values for these phases. 

\begin{figure}[t]
\begin{center}
\includegraphics[height=75mm]{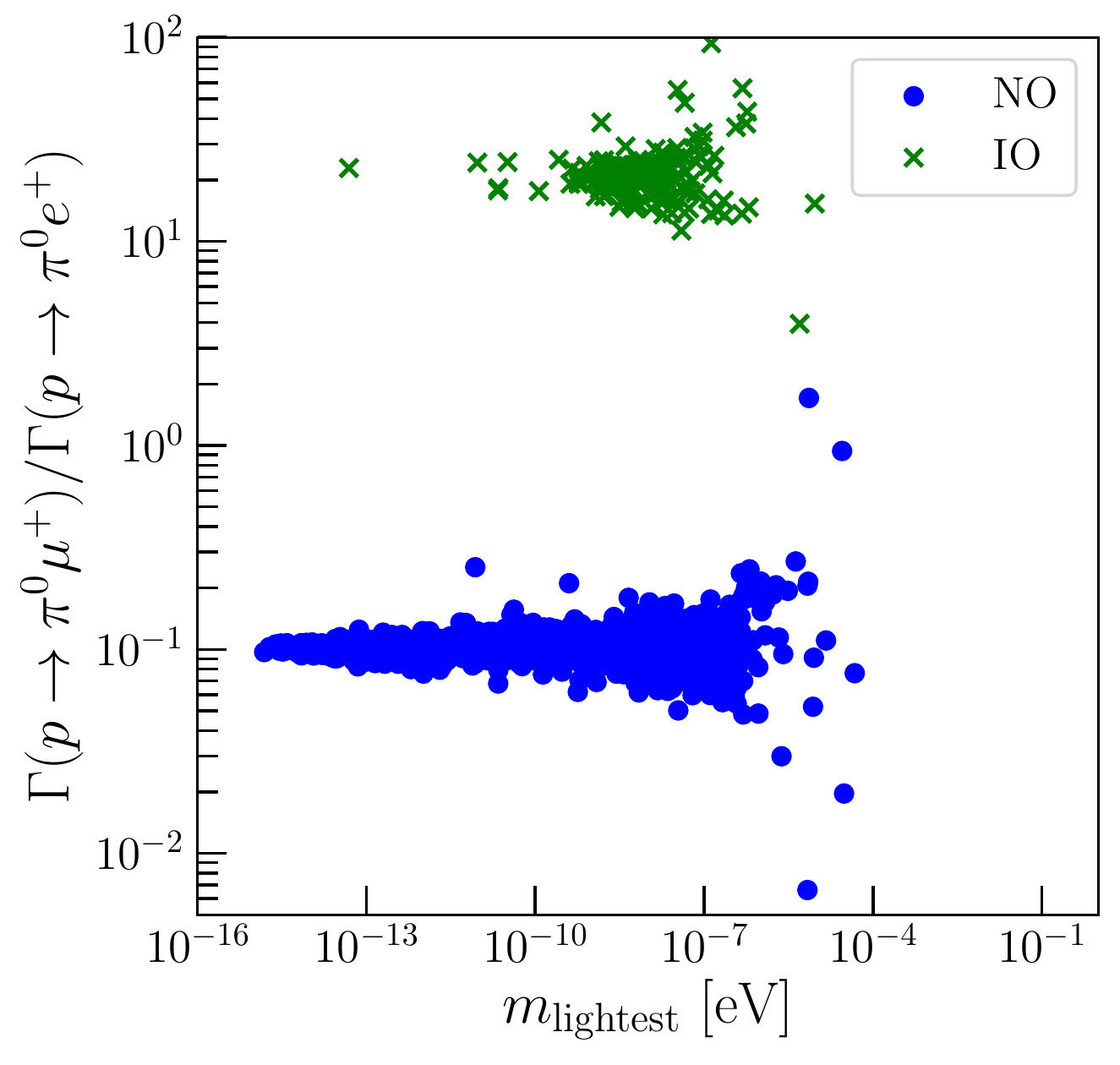}
\caption{\it Scatter plots
of values of $\Gamma (p\to \pi^0 \mu^+)/\Gamma (p\to
 \pi^0 e^+)$ in the flipped SU(5) GUT model for the NO and IO cases
 (blue and green, respectively), 
 as functions of the lightest neutrino mass.}
\label{fig:massdependence}
\end{center}
\end{figure}

The values of $\Gamma (p\to \pi^0 \mu^+)/\Gamma (p\to \pi^0 e^+)$
predicted in the NO and IO flipped SU(5) scenarios are rather
insensitive to the mass of the lightest neutrino, as seen in
Fig.~\ref{fig:massdependence}. On the other hand, we also see there
that the spread in predicted values increases with the lightest
neutrino mass. It may be challenging for the envisaged
next-generation neutrino experiments to measure any deviation from
the central values of the model predictions, but the NO and IO predictions remain well separated and hence distinguishable. 

The predicted values of $\Gamma (p\to \pi^0 \mu^+)/\Gamma
(p\to \pi^0 e^+)$ in flipped SU(5) are much larger than the standard unflipped
SU(5) prediction, which is $\simeq 0.008$. We may therefore be able to
distinguish these two models in future proton decay experiments by
measuring the partial lifetimes of these two decay modes. We can also
determine the neutrino mass ordering in the case of flipped SU(5). 
Proton decay experiments are relatively sensitive to both of these decay
modes, leading to the strongest available constraints on proton partial lifetimes: 
the current limit on $\tau(p\to \pi^0 e^+)$ from
Super-Kamiokande is $2.4 \times 10^{34}$~yrs and that on $\tau(p\to
 \pi^0 \mu^+)$ is $1.6 \times
10^{34}$~yrs \cite{Takenaka, Miura:2016krn} which can be compared to the predicted partial lifetimes given in Eq. (\ref{tp0ef}) and (\ref{tp0mf}), respectively. This makes the ratio $\Gamma
(p\to \pi^0 \mu^+)/\Gamma (p\to \pi^0 e^+)$ given in Eq. (\ref{eq:ptopimuovefl}) interesting for
testing the prediction of flipped SU(5) in future proton decay
experiments such as Hyper-Kamiokande \cite{HKTDR}.

\subsection{$\sum_i \Gamma (p\to \pi^+ \bar{\nu}_i)/\Gamma (p\to \pi^0 e^+)$}

Next we consider the ratio $\sum_i \Gamma (p\to \pi^+
\bar{\nu}_i)/\Gamma (p\to \pi^0 e^+)$. Eqs.~\eqref{eq:ptopinufl}
and \eqref{eq:ptopiefl} imply that for the flipped SU(5)  we have
\begin{align}
 \frac{\sum_{i} \Gamma (p\to  \pi^+ \bar{\nu}_i)_{\rm
 flipped}}{\Gamma (p\to  \pi^0 e^+)_{\rm flipped}}
=
\frac{\left(\langle \pi^+\vert (ud)_Rd_L\vert p\rangle\right)^2 
 }{
\left(\langle \pi^0\vert (ud)_Ru_L\vert p\rangle_e\right)^2
} \frac{1}{|V_{ud}|^2 \left|\left(U_l\right)_{11}\right|^2}
  ~,
\end{align}
whereas for unflipped SU(5) we can use Eqs.~\eqref{eq:ptopinu} and
\eqref{eq:ptopie} to obtain
\begin{align}
\frac{\sum_{i} \Gamma (p\to  \pi^+ \bar{\nu}_i )_{\rm unflipped}}{\Gamma
 (p\to  \pi^0 e^+)_{\rm unflipped}}
= \frac{\left(\langle \pi^+\vert (ud)_Rd_L\vert p\rangle\right)^2 
 }{
\left(\langle \pi^0\vert (ud)_Ru_L\vert p\rangle_e\right)^2
} 
\frac{
 R_A^2 \left|V_{ud}^{} \right|^2 }{
\bigl[
R_A^2 + (1+|V_{ud}|^2)^2 
\bigr]
} 
~.
\end{align}
Setting $R_A = 1$ again, we find 
\begin{equation}
 \frac{\sum_{i} \Gamma (p\to  \pi^+ \bar{\nu}_i )_{\rm unflipped}}{\Gamma
 (p\to  \pi^0 e^+)_{\rm unflipped}} \simeq 0.4 ~.
\label{eq:rptopinusu5}
\end{equation}
We note, however, that in the supersymmetric standard SU(5)
GUT colour-triplet Higgs
exchange also induces $p \to \pi^+ \bar{\nu}$ (see, for instance,
Refs.~\cite{Nagata:2013sba, Nagata:2013ive, eemno, Ellis:2019fwf}),
which can be much larger than the contribution in
Eq.~\eqref{eq:ptopinu}. Therefore, the value in
Eq.~\eqref{eq:rptopinusu5} should be regarded as a lower limit on
$\sum_i \Gamma (p\to \pi^+ \bar{\nu}_i)/\Gamma (p\to \pi^0 e^+)$ in
standard unflipped SU(5).

\begin{figure}[!ht]
\begin{center}
\includegraphics[height=75mm]{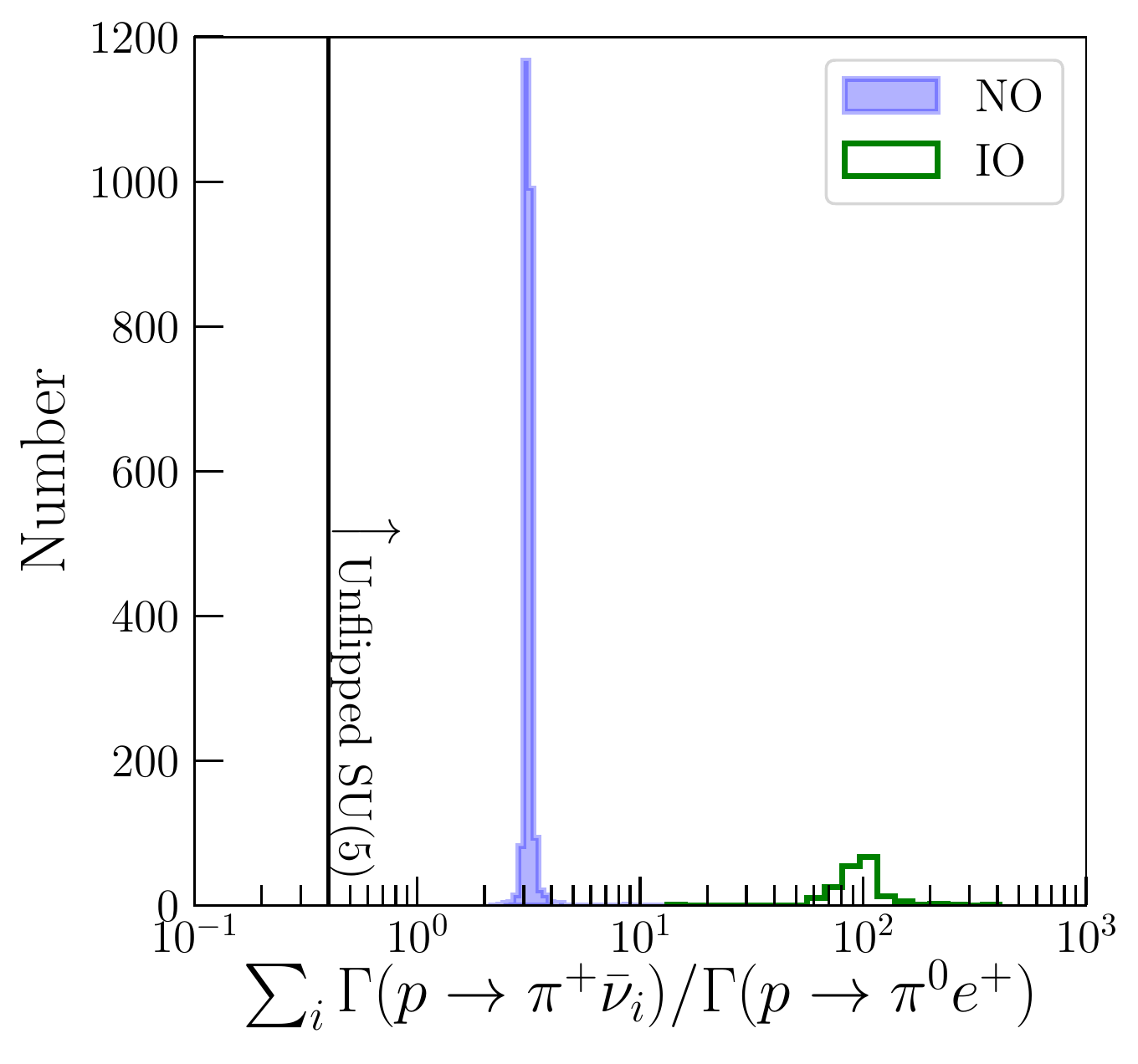}
\caption{\it Histograms of $\sum_{i}\Gamma (p\to \pi^+ \bar{\nu}_i)/\Gamma (p\to
 \pi^0 e^+)$ in flipped SU(5) for the NO and IO cases in blue and
 green, respectively. The unflipped SU(5) prediction has a lower limit
 shown as the vertical solid line. }
\label{fig:rpinuovpie}
\end{center}
\end{figure}

We show in Fig.~\ref{fig:rpinuovpie} histograms of $\sum_{i}\Gamma (p\to
\pi^+ \bar{\nu}_i)/\Gamma (p\to \pi^0 e^+)$ in the flipped SU(5) model for the
NO and IO cases in blue and green, respectively. Unflipped SU(5) has
the lower limit indicated by the vertical solid line. As in
the previous subsection, we can again estimate this ratio using the
approximation given in Eq.~\eqref{eq:ul11}: 
\begin{align}
 \frac{\sum_{i} \Gamma (p\to  \pi^+ \bar{\nu}_i)_{\rm
 flipped}}{\Gamma (p\to  \pi^0 e^+)_{\rm flipped}}
=
\frac{\left(\langle \pi^+\vert (ud)_Rd_L\vert p\rangle\right)^2 
 }{
\left(\langle \pi^0\vert (ud)_Ru_L\vert p\rangle_e\right)^2
} \frac{1}{|V_{ud}|^2 (c_{12} c_{13})^2}
 \simeq 3.15 ~,
\end{align}
for NO, and 
\begin{align}
 \frac{\sum_{i} \Gamma (p\to  \pi^+ \bar{\nu}_i)_{\rm
 flipped}}{\Gamma (p\to  \pi^0 e^+)_{\rm flipped}}
=
\frac{\left(\langle \pi^+\vert (ud)_Rd_L\vert p\rangle\right)^2 
 }{
\left(\langle \pi^0\vert (ud)_Ru_L\vert p\rangle_e\right)^2
} \frac{1}{|V_{ud}|^2 s_{13}^2}
\simeq 94.8  ~,
\end{align}
for IO, which agree with the results shown in
Fig.~\ref{fig:rpinuovpie}. 

This ratio is, however, less powerful for distinguishing the
flipped and unflipped SU(5) GUTs than $\Gamma (p\to \pi^0 \mu^+)/\Gamma (p\to \pi^0
e^+)$. First, due to the potential contribution of the colour-triplet
Higgs exchange, we have only a lower limit on the unflipped SU(5)
prediction. Since the predicted values in the flipped SU(5) are larger
than this lower limit, the unflipped SU(5) prediction can in principle
mimic the flipped SU(5) predictions. Secondly, the sensitivities of
experiments to $p \to \pi^+ \bar{\nu}$ and $n \to \pi^0 \bar{\nu}$ tend to be much worse than that
to $p \to \pi^0 \mu^+$; the present bound on $p \to \pi^+ \bar{\nu}$ from
Super-Kamiokande is $\tau (p \to \pi^+ \bar{\nu}) > 3.9 \times
10^{32}$~yrs and that on $\tau (n \to \pi^0 \bar{\nu}) > 1.1 \times
10^{33}$~yrs~\cite{Abe:2013lua}, which are much lower than the limit on
$p \to \pi^0 \mu^+$. On the other hand, the value of $\sum_{i}\Gamma (p\to
\pi^+ \bar{\nu}_i)/\Gamma (p\to \pi^0 e^+)$ predicted in the flipped SU(5) model 
in the IO case is so large that this might be detectable.

\subsection{$\Gamma (p\to K^0 e^+)/\Gamma (p\to \pi^0 e^+)$}

The ratio $\Gamma (p\to K^0 e^+)/\Gamma (p\to \pi^0 e^+)$ in flipped
SU(5) is computed from Eqs.~\eqref{eq:ptokefl} and \eqref{eq:ptopiefl}
to be
\begin{align}
\frac{\Gamma (p\to  K^0 e^+ )_{\rm flipped}}{\Gamma (p\to  \pi^0
 e^+)_{\rm flipped}}
= 
\frac{(m_p^2 - m_K^2)^2}{(m_p^2 - m_\pi^2)^2}
\frac{\left(\langle K^0 \vert (us)_Ru_L\vert p\rangle_e\right)^2 
 }{
\left(\langle \pi^0\vert (ud)_Ru_L\vert p\rangle_e\right)^2
} 
\frac{
 \left|V_{us}^{} \right|^2 }{
 \left|V_{ud}^{} \right|^2
} 
\simeq 1.8 \times 10^{-2} ~.
\label{eq:rptokeovpiefl}
\end{align}
As we see, this ratio does not depend on the matrix $U_l$. In
unflipped SU(5), we use Eqs.~\eqref{eq:ptoke} and \eqref{eq:ptopie} to
find 
\begin{align}
\frac{\Gamma (p\to  K^0 e^+ )_{\rm unflipped}}{\Gamma (p\to  \pi^0
 e^+)_{\rm unflipped}}
= 
\frac{(m_p^2 - m_K^2)^2}{(m_p^2 - m_\pi^2)^2}
\frac{\left(\langle K^0 \vert (us)_Ru_L\vert p\rangle_e\right)^2 
 }{
\left(\langle \pi^0\vert (ud)_Ru_L\vert p\rangle_e\right)^2
} 
\frac{
 \left|V_{ud}^{} V_{us}^*\right|^2 }{
\bigl[
R_A^2 + (1+|V_{ud}|^2)^2 
\bigr]
} 
\simeq 3.3 \times 10^{-3}
~,
\end{align}
for $R_A = 1$.
The contribution of the colour-triplet Higgs exchange to $p\to  K^0 e^+$
is negligible unless flavour violation occurs in sfermion mass 
matrices~\cite{Nagata:2013sba, Nagata:2013ive}, so this value can be
regarded as a prediction of unflipped SU(5). As we see, this
unflipped SU(5) prediction is much lower than the flipped SU(5)
prediction \eqref{eq:rptokeovpiefl}, and thus we can in principle also use
the ratio $\Gamma (p\to K^0 e^+)/\Gamma (p\to \pi^0 e^+)$ to distinguish
between these two GUT models.

\subsection{$\Gamma (p\to K^0 \mu^+)/\Gamma (p\to \pi^0 \mu^+)$}

From Eqs.~\eqref{eq:ptokmufl} and \eqref{eq:ptopimufl}, we have 
\begin{align}
\frac{\Gamma (p\to  K^0 \mu^+ )_{\rm flipped}}{\Gamma (p\to  \pi^0
 \mu^+)_{\rm flipped}}
= 
\frac{(m_p^2 - m_K^2)^2}{(m_p^2 - m_\pi^2)^2}
\frac{\left(\langle K^0 \vert (us)_Ru_L\vert p\rangle_\mu\right)^2 
 }{
\left(\langle \pi^0\vert (ud)_Ru_L\vert p\rangle_\mu\right)^2
} 
\frac{
 \left|V_{us}^{} \right|^2 }{
 \left|V_{ud}^{} \right|^2
} 
\simeq 0.02
~.
\end{align}
Again, this ratio does not depend on the matrix $U_l$. In unflipped
SU(5), Eqs.~\eqref{eq:ptokmu} and \eqref{eq:ptopimu} lead to 
\begin{align}
\frac{\Gamma (p\to  K^0 \mu^+ )_{\rm unflipped}}{\Gamma (p\to  \pi^0 \mu^+)_{\rm unflipped}}
= 
\frac{(m_p^2 - m_K^2)^2}{(m_p^2 - m_\pi^2)^2}
\frac{\left(\langle K^0 \vert (us)_Ru_L\vert p\rangle_\mu\right)^2 
 }{
\left(\langle \pi^0\vert (ud)_Ru_L\vert p\rangle_\mu\right)^2
} 
\frac{
\bigl[
R_A^2 + (1+|V_{us}|^2)^2 
\bigr]
} {
 \left|V_{ud}^{} V_{us}^*\right|^2 }
\simeq 16.7
~,
\end{align}
for $R_A = 1$. The contribution of colour-triplet Higgs exchange to $p\to  K^0 \mu^+$
is small unless flavour violation occurs in sfermion mass 
matrices~\cite{Nagata:2013sba, Nagata:2013ive}. Therefore, this ratio can again
be used to distinguish between the flipped and unflipped SU(5) GUTs.

\subsection{$p \to K^+ \bar{\nu}$}

This process tends to be the dominant decay mode in the supersymmetric
standard unflipped SU(5) GUT model~\cite{Sakai:1981pk}. In flipped SU(5), on the
other hand, as seen in Eq.~\eqref{eq:ptoknufl}, we have \cite{Ellis:1993ks}
\begin{align}
 \Gamma (p\to  K^+ \bar{\nu}_i)&=
0 ~.
\end{align}
This is a distinctive prediction in flipped SU(5)---if this decay mode
is discovered in future proton decay experiments, flipped SU(5) is
excluded.

\section{Discussion and Prospects}
\label{sec:conclusion}

\begin{figure}[!ht]
\begin{center}
\includegraphics[height=55mm]{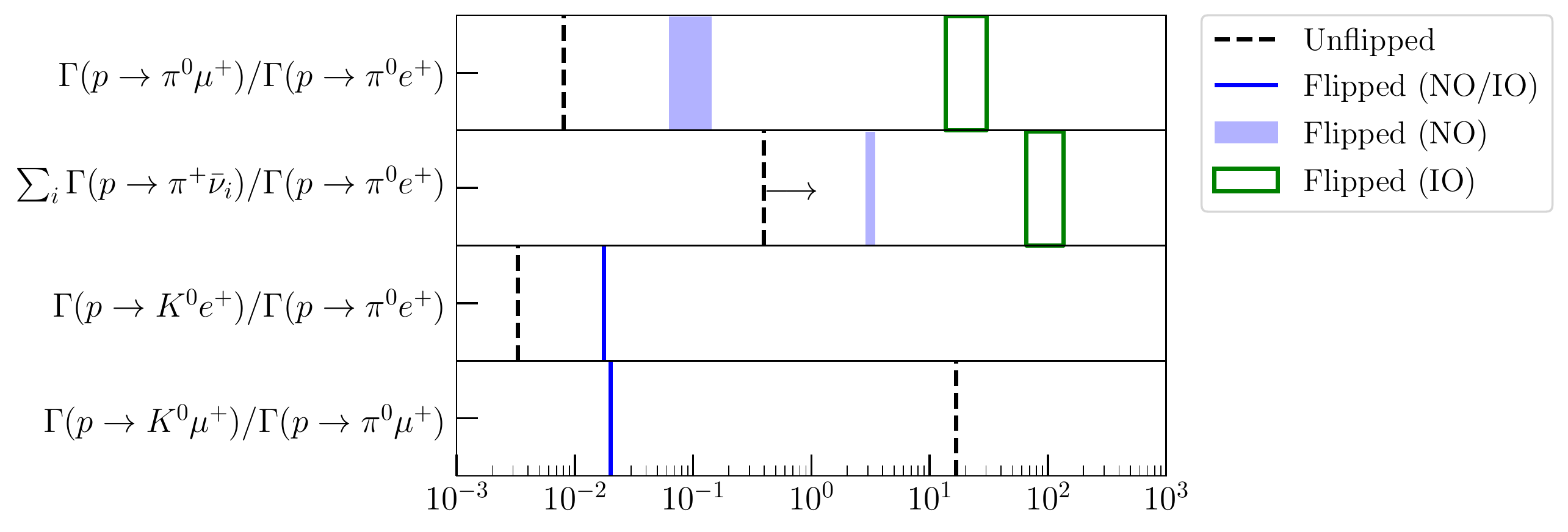}
\caption{\it Compilation of the ratios of proton decay rates predicted in standard unflipped SU(5) (dashed black lines), no-scale flipped SU(5) with neutrino masses that are normal-ordered (NO, blue shading) or inverse-ordered (IO, green boxes). Cases where the flipped SU(5) predictions are independent of the neutrino mass ordering are indicated by solid blue lines.}
\label{fig:prediction}
\end{center}
\end{figure}

We have explored in this paper various nucleon decay modes in the
flipped SU(5) GUT model developed in~\cite{egnno2, egnno3, egnno4, egnno5},
which builds upon earlier studies~\cite{Barr, DKN, flipped2, AEHN, Ellis:1993ks}.
We have presented flipped SU(5) predictions in scenarios with both normal-ordered
neutrino masses (NO) and inverse ordering (IO), and compared them with the
predictions of the standard unflipped SU(5) GUT. Our results for the ratios of decay rates
$\Gamma(p \to \pi^0 \mu^+)/\Gamma(p \to \pi^0 e^+)$, 
$\Gamma(p \to \pi^+ \bar \nu)/\Gamma(p \to \pi^0 e^+)$,
$\Gamma(p \to K^0 e^+)/\Gamma(p \to \pi^0 e^+)$ and
$\Gamma(p \to K^0 \mu^+)/\Gamma(p \to \pi^0 \mu^+)$
are compiled in Fig.~\ref{fig:prediction}.~\footnote{We note also the flipped 
SU(5) prediction that $\Gamma(p \to K^+ \bar{\nu})$ vanishes.}
In all cases we see clear differences between the predictions
of flipped SU(5) and standard SU(5), and in the cases of
$\Gamma(p \to \pi^0 \mu^+)/\Gamma(p \to \pi^0 e^+)$ and 
$\Gamma(p \to \pi^+ \bar \nu)/\Gamma(p \to \pi^0 e^+)$
we also see clear distinctions between the NO and IO
predictions. 

The `Golden Ratio' from the point of view of our analysis is
$\Gamma(p \to \pi^0 \mu^+)/\Gamma(p \to \pi^0 e^+)$. We recall that 
Super-Kamiokande has similar sensitivities to these two decay modes,
and has established limits on their partial lifetimes of $1.6$ and $2.4 \times 10^{34}$~yrs, respectively~\cite{Takenaka, Miura:2016krn}.~\footnote{The corresponding searches for $n \to \pi^- l^+$ are less constraining; 
the present limits on the lifetimes of these decay modes, $\tau (n \to \pi^- e^+) > 5.3 \times 10^{33}$~yrs and $\tau (n \to \pi^- \mu^+) > 3.5 \times 10^{33}$~yrs \cite{TheSuper-Kamiokande:2017tit}, are weaker than those on $p \to \pi^0 l^+$, while the predicted partial decay widths are larger, as shown in Eq.\eqref{npi-l+}.} We expect that
future proton decay experiments such as Hyper-Kamiokande \cite{HKTDR}
should have an order of magnitude greater sensitivity to both these
decay modes, and hence have a window of opportunity to probe both
the NO and IO predictions. Indeed, in the IO case a search for $p \to \pi^0 \mu^+$
with a sensitivity to a partial lifetime of $10^{35}$~yrs would constrain the model
as much as a sensitivity to $p \to \pi^0 e^+$ of about $2 \times 10^{36}$~yrs.

Our results highlight the importance of targeting proton decay
modes involving final-state particles from different generations,
since our `Golden Ratio' and two others, 
$\Gamma(p \to K^0 e^+)/\Gamma(p \to \pi^0 e^+)$ and
$\Gamma(p \to K^0 \mu^+)/\Gamma(p \to \pi^0 \mu^+)$, involve mixtures of
first- and second-generation leptons and quarks. 

The fourth ratio,
$\Gamma(p \to \pi^+ \bar \nu)/\Gamma(p \to \pi^0 e^+)$,
does not involve identifiable second-generation fermions, but
second- and third-generation neutrinos contribute to the enhanced
values of the ratio predicted in the two flipped SU(5) scenarios
we have studied. The current limit on $\tau(p \to \pi^+ \bar{\nu})$ is only $3.9 \times 10^{32}$~yrs~\cite{Abe:2013lua}.
However, in the IO model this lifetime would be two orders of magnitude shorter than
$\tau(p \to \pi^0 e^+)$, so the current limit corresponds to $\tau (p \to \pi^0 e^+) > 3.7 \times 10^{34}$~yrs. 
Hence the search for $p \to \pi^+ \bar{\nu}$ currently sets a tighter constraint on the IO model than
that set by the $p \to \pi^0 e^+$ search. We are unaware of estimates of the improved sensitivity to 
$p \to \pi^+ \bar \nu$ of the upcoming large neutrino experiments, but increasing the sensitivity to
$p \to \pi^+ \bar \nu$ by the same factor as anticipated for $p \to \pi^0 e^+$~\cite{HKTDR} would 
constrain the IO model as much as a sensitivity to the latter mode of $> 3 \times 10^{35}$~yrs. The current limit $\tau(n \to \pi^0 \bar \nu) > 1.1 \times 10^{33}$~yrs \cite{Abe:2013lua} constrains the IO model even more, since it corresponds to $\tau (p \to \pi^0 e^+) > 5 \times 10^{34}$~yrs. Again, we are unaware of any estimate of the sensitivity in a future experiment, but an order-of-magnitude improvement would correspond to $\tau (p \to \pi^0 e^+) > 5 \times 10^{35}$~yrs.

These examples show that if the upcoming large neutrino experiments do discover nucleon decay,
they will have interesting opportunities to explore both GUT and flavour physics.\\
~~\\
\section*{Acknowledgements}
The work of J.E. was supported partly by the United Kingdom STFC Grant
ST/P000258/1 and partly by the Estonian Research Council via a Mobilitas
Pluss grant. The work of M.A.G.G. was supported by the Spanish Agencia
Estatal de Investigaci\'on through the grants FPA2015-65929-P
(MINECO/FEDER, UE), PGC2018095161-B-I00, IFT Centro de Excelencia Severo
Ochoa SEV-2016-0597, and Red Consolider MultiDark
FPA2017-90566-REDC. The work of N.N. was supported by the Grant-in-Aid
for Young Scientists B (No.17K14270) and Innovative Areas
(No.18H05542). N.N. would also like to thank the members of
William I. Fine Theoretical Physics Institute for their hospitality and
financial support while finishing this work. 
The work of D.V.N. was supported partly by the DOE grant
DE-FG02-13ER42020 and partly by the Alexander S. Onassis Public Benefit
Foundation. The work of K.A.O. was supported partly by the DOE grant
DE-SC0011842 at the University of Minnesota.



\end{document}